\documentclass[ %
 reprint,
 amsmath,amssymb,
 aps,
prb,
]{revtex4-2}
\usepackage{xcolor} 
\usepackage[normalem]{ulem} 
\usepackage{graphicx}
\usepackage{dcolumn}
\usepackage{bm}
\usepackage{hyperref}
\usepackage[mathlines]{lineno}

\begin{document}

\preprint{APS/123-QED}

\title{Relativistic Dynamics and Electron Transport in Isolated Chiral Molecules}

\author{Sushant Kumar Behera}
\thanks{Equally contributed to this work.}%
\affiliation{Department of Physics, Università degli Studi di Pavia, Via Bassi 6, 27100 Pavia, Italy.}
\author{Ruggero Sala}
\thanks{Equally contributed to this work.}%
\affiliation{Department of Chemistry, Università degli Studi di Pavia, Via Taramelli 12, 27100 Pavia, Italy.}
\affiliation{University School for Advanced Studies (IUSS), Piazza della Vittoria 15, 27100 Pavia, Italy.}
\author{Abhirup Roy Karmakar}
\affiliation{Department of Chemistry, Università degli Studi di Milano, Via Golgi 19, 20133 Milano, Italy.}
\author{Matteo Moioli}
\affiliation{Department of Physics, Università degli Studi di Pavia, Via Bassi 6, 27100 Pavia, Italy.}
\author{Rocco Martinazzo}
\email{rocco.martinazzo@unimi.it}
\affiliation{Department of Chemistry, Università degli Studi di Milano, Via Golgi 19, 20133 Milano, Italy.}
\author{Matteo Cococcioni}
\email{matteo.cococcioni@unipv.it}
\affiliation{Department of Physics, Università degli Studi di Pavia, Via Bassi 6, 27100 Pavia, Italy.}

\date{\today}

\begin{abstract}
The Chirality-Induced Spin Selectivity (CISS) effect describes the ability of chiral molecules and crystals to transmit spin-polarized currents, a phenomenon first identified in 1999. Although this effect holds great promise for a broad spectrum of different applications in device physics and synthetic chemistry (including, e.g., spintronics, quantum computing, spin- and enantio-selective chemistry), its underlying mechanisms remain incompletely understood. The prevailing hypothesis attributes the CISS effect to enhanced spin-orbit coupling (SOC) within chiral molecules. However, the SOC magnitude required to align with experimental observations significantly exceeds the values derived from conventional atomic-scale calculations, particularly for systems composed of light atoms. In this work, we leverage the implementation of \texttt{fully relativistic density functional theory (DFT)} equations, as available in the \texttt{Dirac code}, to investigate how molecular chirality manifests itself in the chirality density of electronic states. We further explore how this responds to an applied external electric field. To assess spin-dependent transport, we employ the \texttt{Landauer-Imry-Büttiker} formalism, examining the dependence of spin transmission on the twist angle of the molecular structure that defines its geometrical chirality. While our findings qualitatively align with experimental trends, they point to the necessity of a more general treatment of SOC, \textit{e.g.}, including geometrical terms or through the dependence of advanced exchange-correlation functionals on the electronic spin-current density.
\end{abstract}

\maketitle
\section{Introduction}

Chirality-Induced Spin Selectivity (CISS) is a transformative discovery in spintronics, as it enables the spin polarization of charge carriers without requiring external magnetic fields \cite{Brian2024}. First observed in 1999 \cite{Kaushik1999, Naaman2012}, CISS occurs when electrons traverse chiral systems \cite{Brian2024}, such as helical molecules  \cite{Anup2017} or noncentrosymmetric materials \cite{Naaman2019}, resulting in currents that exhibit a significant spin polarization (in some cases close to 100$\%$) \cite{Desmarais2024}. This phenomenon has far-reaching implications in spintronics \cite{Naaman2015}, redox chemistry \cite{Anupkumar2017, Karen2016}, particularly for spin-selective electron transport \cite{Lou2007} and quantum computing \cite{Wang2024, Hannah2023}. The mechanisms driving CISS remain an active area of research. Traditionally, spin-orbit coupling (SOC) was linked to heavy atoms with strong nuclear potentials and its effects were expected to be relevant in the region around these nuclei \cite{Jose2018, Marcin2019}. Recent studies suggest that local electric fields induced by chirality \cite{Guinea2010, Abdiel2024}, perpendicular to electron motion, may enhance SOC far from atomic nuclei, significantly influencing spin dynamics and transport in materials like chiral nanowires \cite{Fan2020}, molecular magnets \cite{Yang2021}, and curved graphene that do not necessarily contain heavy atoms \cite{Levy2010}.

CISS is broadly rooted in the interplay of chirality and electron spin, arising from broken inversion symmetry in chiral systems \cite{Bing2023}. This symmetry breaking causes asymmetric transmission of \textit{spin-up} and \textit{spin-down} electrons, producing pronounced spin polarization. Initially observed in biomolecules like DNA and peptides \cite{Gohler2011}, CISS has since expanded to synthetic systems, highlighting its versatility \cite{Yang2015}. Definitions of CISS vary, from asymmetric scattering of spin-polarized electrons to selective interactions with magnetized surfaces of chiral species or spin polarization in chiral systems \cite{Koyel2018}. These interpretations share a focus on spin and structural chirality but may describe distinct phenomena, complicating the perspective of reaching a unified theory. Theoretical studies of the CISS largely focus on transport phenomena, some of which highlight that spin-resolved transmission is related to spin current, not the charge current observed in experiments. Magnetocurrent requires intrinsic spin anisotropy, often introduced artificially, and spin current studies may offer insights into CISS origins \cite{Spivak2004,Huisman2023}. Using Green’s functions and the Landauer-Büttiker formula \cite{Buttiker1993}, models link spin-dependent transmission to currents, employing methods like inelastic scattering to break Onsager reciprocity \cite{Luo2020}. Meanwhile, non-equilibrium Green's function (NEGF) \cite{You2000} transport calculations show that SOC introduces spin-dependence in transmission, preserving Hamiltonian hermiticity and reflecting true chirality, where systems are parity-odd (\textit{P-odd}) and time-reversal-even (\textit{T-even}) \cite{Zollner2020, Dednam2023}. Meanwhile, experiments confirm CISS at the single-molecule level, even with nonlocal effects. High spin polarization was observed in helicenes \cite{Kiran2016}, molecular knots \cite{Zhang2023}, and supramolecular assemblies \cite{Kulkarni2020}, with symmetry breaking arising from non-covalent bonding. Metal-organic frameworks with Dy atoms achieved nearly 100$\%$ spin polarization, highlighting SOC's role but potentially masking ligand-based CISS effects \cite{Uxua2020}. For now, CISS is best viewed as a collection of related effects with promising applications in quantum computing \cite{Grinolds2011}, superconductivity \cite{Linder2015}, and optoelectronics \cite{Long2020}.

In order to reach a comprehensive understanding about the mechanism of CISS, it is convenient to refer to the four-component (\textit{4C}) relativistic Dirac wave equation from which SOC ultimately originates. In the Dirac equation, the free particle spinor wave function with a positive-definite probability density is given by: \[(i\gamma^\mu \partial_\mu - m)\psi = 0,\] where \(\mu = 0,1,2,3\) represents the space-time indices, and \(\gamma^\mu\) are the \textit{\(4 \times 4\)} gamma matrices. The field \(\psi\) is now a \textit{4C} Dirac spinor. Aside from the four basis gamma matrices, a fifth one has been defined as the product of the others: \[\gamma^5 \equiv i\gamma^0 \gamma^1 \gamma^2 \gamma^3.\]Here, the \(\gamma^5\) matrix defines fermions' chirality and is, in fact, diagonal in the Weyl basis, enabling the separation of chiral components \cite{Matsuo2011, Shitade2020}. Studying the chirality density defined by $\gamma^5$ could thus enhance understanding of spin-related phenomena like CISS, especially under the most general conditions. Using the relativistic approach based on \texttt{4C Dirac-equation} and relying on the SOC it implicitly contains, we study how the structural chirality of a molecule reflects itself on the chirality of its electronic states, and how the distribution of this quantity is changed by an applied external electric field. Moreover, spin-dependent transmission studies have been conducted in dependence of the twist angle of the structure that measures the geometrical chirality. While qualitatively consistent with experimental observations, our results point to the need for more advanced exchange-correlation functionals that, based \textit{e.g.} on a dependence on the spin-current density \cite{Desmarais2024} are able to capture the effect of a more general definition of SOC.

\section{Methodology}
\subsection{Relativistic origin of SOC and Chirality in the Dirac equation}
We performed \texttt{ab initio} calculations using the \texttt{Dirac24} \cite{Saue2020} relativistic quantum chemistry code on a set of small, isolated molecular system. The spinor wavefunction was optimized self-consistently, followed by the computation of expectation values using the pre-existing routines implemented in this code. The calculations employed the relativistic density functional theory (DFT) framework, specifically utilizing the \textit{Dirac-Kohn-Sham (DKS)} Hamiltonian as outlined in previous works \cite{Saue2011, Saue2002, Daniel2020}. The \(\gamma^5\) matrix projects out the left- and right-handed components of the spinor wavefunction, directly linking to chirality in relativistic quantum mechanics. The chirality density at any spacetime point \(x\) can be defined as: \[\chi(x) = \sum_{\mu, \nu} \psi_\mu^*(x)\gamma_\mu \gamma^5 \psi_\nu(x)\] 

In the weak relativistic limit, the 4C Dirac spinors can be mapped into two-component (\textit{2C}) Pauli spinors \((\phi)\), with spin-up (\(\alpha\)) and spin-down (\(\beta\)) components. In this limit, chirality density aligns with the helicity (\textit{i.e.}, the projection of the electron's spin onto its linear momentum), linking particle chirality to \texttt{spin-momentum locking (SML)} alignment \cite{Krieger2024} and spin-polarized currents in CISS effects.

A brief evaluation of the effects of different basis sets and functionals on the computed quantities of interest showed comparable results. Consequently, all calculations presented here were conducted with the \textit{dyall.3zp} triple-zeta quality basis set \cite{Dyall1994}, which includes polarization functions, in combination with the exchange correlation functional \textit{local density approximation} (LDA) \cite{Lee1994}. To ensure proper symmetry handling in the four-component calculations, the relativistic basis set was kept uncontracted throughout. The isolated molecule of the present model was selected based on three criteria: (1) it should consist of light atoms to minimize strong atomic spin-orbit coupling (SOC) effects, keeping the system closer to experimental relevance; (2) it should allow tunable chirality, enabling both chiral and achiral configurations through simple variation of a geometric parameter; (3) it should be small enough to ensure reasonable computational cost.

Here, we consider ethane ($CH_3-CH_3$) as the isolated molecule, where twisting the $CH_3$ groups around the $C-C$ bond (defined by the $H-C-C-H$ dihedral angle \textit{d}) produces achiral structures for $d=0^{\circ}$ and $d=\pm30^{\circ}$, and chiral enantiomers for $0^{\circ}<d<30^{\circ}$ and $-30^{\circ}<d<0^{\circ}$. The \textit{z}-axis was defined along the $C-C$ bond, with \textit{x} and \textit{y} axes equivalent by symmetry. The eclipsed ($d=0^{\circ}$), staggered ($d=\pm30^{\circ}$), and skew ($d~\neq~0^{\circ},~\pm30^{\circ}$) conformations correspond to point groups $D_{3h}$, $D_{3d}$, and $D_{3}$, respectively. To explore effects of molecular dipole moments, \textit{1,1,1-trichloroethane} ($CCl_3-CH_3$) was also considered, introducing a \textit{z}-axis dipole aligned from $CCl_3$ to $CH_3$ group. Its analogous configurations yield $C_{3v}$, $C_{3v}$, and $C_{3}$ symmetry groups. Finally, \textit{polyphenyl} systems, such as triphenyl molecules, were explored for their tunable chemical structures and potential to study the dependence of length in CISS, as well as its suitability for future experimental investigations. The dihedral angles of $0^{\circ}$ and $90^{\circ}$ between adjacent benzene rings correspond to the symmetries of $D_{2h}$, while intermediate angles produce chiral structures in the group $D_{2}$. Molecular visualizations were conducted using the \textit{Avogadro} software \cite{Hanwell2012}.

\subsection{Electron Transport with Spin-Orbit Coupling}
\subsubsection{Landauer-Imry-Büttiker formalism}
\begin{figure}
\centering
\includegraphics[width=7.0cm,height=6.0cm]{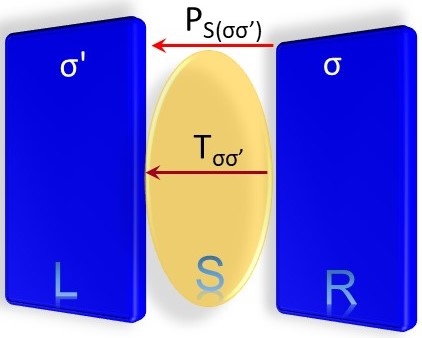}
 \caption{Sketch of the left ($L$) and right ($R$) two-terminal lead setup for transport study. Here, spin polarization, $P_S$ has been calculated as a function of energy ($E$) with transmission function $T$. For electrons, the tunneling current originating with right spin orientation \(\sigma_R\) to left spin orientation \(\sigma'_L\) with a functional relation \((\sigma, \sigma' \in \{\uparrow, \downarrow\})\) can be calculated.}
  \label{scheme}
\end{figure}

CISS is a phenomenon observed for charge carriers in motion (currents). Therefore, for a proper account of its entity it is mandatory to perform transport calculations. Here we apply the simplest possible scheme, within the Landauer approach. In this scheme, the SOC contributions captured by the DFT Hamiltonian matrix effectively describe spin polarization in closed-shell systems and qualitatively align with CISS behavior under chirality changes. Thus, we apply the Landauer-Imry-Büttiker (LIB) model \cite{Neese2005,Sepunaru2015}, which simulates transport through coherent tunneling. This method is ideal for isolated molecular systems, especially when one wants to focus on ballistic regimes and is not much interested in scattering. The LIB approach, integrated with the DFT first-principles, employs a partitioning scheme \cite{Caroli1971} that divides the junction Hamiltonian into three regions: left lead (L), scattering region (S), and right lead (R) (shown in Fig. \ref{scheme}). This requires single-particle basis functions localized to their respective regions, ensuring an effective description of transport phenomena. The studied system exhibits notable spin polarization ($P_S$), likely influenced by the SOC from the electrodes in the junction. Here, this $P_S$ value is a crucial property for CISS. Conventional DFT primarily accounts for one-electron SOC terms, often neglecting many-body contributions to SOC that could significantly impact the CISS effect. Including these SOC terms may enhance the intrinsic SOC of the chiral molecule, inducing spin polarization even without metal electrodes. This in fact was the original motivation for using a fully relativistic description of the electronic states of the system using the Dirac code. Here, the transmission functions (\textit{T}) were computed, which incorporates the SOC terms. Results suggest that the SOC terms amplify the intrinsic SOC of chiral systems.

\section{Results and Discussions}
\subsection{Static calculations for isolated molecular Systems}
\subsubsection{Isolated molecular geometry}
\begin{figure}
\centering
\includegraphics[width=8.5cm,height=7.0cm]{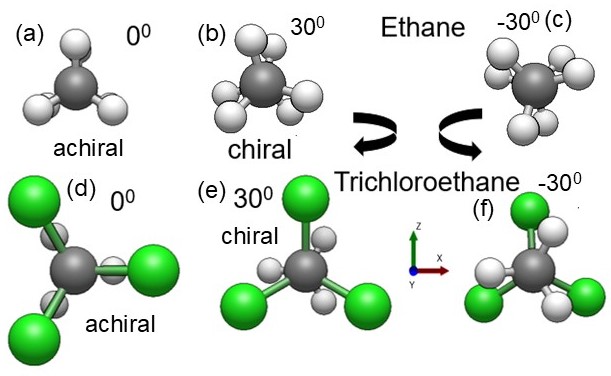}
 \caption{Geometrical chirality of the isolated molecular structure of ethane (a) $0^{\circ}$ eclipsed, achiral $D_{3h}$ to (b) skewed $30^{\circ}$ and (c) $-30^{\circ}$. Similarly, trichloroethane shows (d) $0^{\circ}$ achiral, (e) skew configurations ($30^{\circ}$, $C_3$)and (f) $-30^{\circ}$ in chiral forms viewed along \textit{z}-axis.}
  \label{figure01}
\end{figure}

The first system we examine here is ethane (C$H_3$–C$H_3$) as an isolated molecule (\textit{refer to the top panel of} Fig. \ref{figure01} (a-c), where rotation of the methyl (C$H_3$) groups about the C–C bond, defined by the H–C–C–H dihedral angle (\textit{d}), results in distinct structural symmetries. Achiral configurations occur at \textit{d} = $0^\circ$ (eclipsed) and \textit{d} = $\pm 30^\circ$ (staggered), while chiral enantiomers arise for $0^\circ < \textit{d} < 30^\circ$ and $-30^\circ < \textit{d} < 0^\circ$. The molecular \textit{zz}-axis is aligned along the C–C bond, with equivalent \textit{xx} and \textit{yy} axes due to symmetry. The eclipsed, staggered, and skewed (\textit{d} $\neq$ $0^\circ$, $\pm 30^\circ$) conformations correspond to the point groups $D_{3h}$, $D_{3d}$, and $D_{3}$, respectively. To investigate the effects of molecular dipole moment, we also analyzed \textit{1,1,1-trichloroethane} (C$H_3$-C$Cl_3$), which introduces a dipole moment along the \textit{zz}-axis, directed from the chlorine (Cl) atoms toward the hydrogen (H) atom (\textit{refer to the bottom panel of} Fig. \ref{figure01} (d-f)). Its analogous conformations exhibit $C_{3v}$ symmetry for both the eclipsed and staggered configurations, while the skewed conformation corresponds to the $C_{3}$ point group.

\subsection{Static calculations without electric field}
\subsubsection{Chirality density}
\begin{figure}
\centering
\includegraphics[width=8.5cm,height=6.5cm]{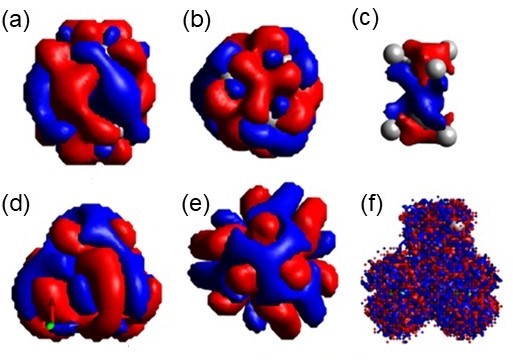}
 \caption{(a) Side view of the chiral $-30^{\circ}$ ethane molecule, showing accumulated chirality density with the same sign (red for positive, blue for negative, isovalue = \(1 \times 10^{-10}\)) at each end of the molecule. (b) View along the \(z\)-axis, showing that the chirality density respects the \(C_3\) axis symmetry. (c) Same molecule with an increased isovalue of \(1 \times 10^{-9}\), matching the isovalue used for the trichloroethane molecules (d-f) for comparison. (d) Side view of the chiral $+30^{\circ}$ trichloroethane molecule, showing anisotropic accumulation of chirality density along the \(z\)-axis due to broken symmetry and the molecule's intrinsic dipole. (e) View along the \(z\)-axis, showing higher-order multipolar terms in the chirality density distribution. (f) Achiral eclipsed $0^{\circ}$ trichloro-ethane molecule, with randomly distributed chirality density (isovalue = \(1 \times 10^{-21}\)).
}
  \label{figure02}
\end{figure}

Building on the work of Hoshino et al. \cite{Hoshino2023}, who defines chirality density as an intrinsic property of localized orbitals, it can be argued that CISS phenomenon corresponds to a multipolar interaction that becomes active exclusively under chiral point group symmetries. To explore this concept, we calculated the chirality density for various dihedral angles in both ethane and \textit{1,1,1-trichloroethane} molecular systems. In achiral configurations, the chirality density manifests as numerical noise, with its integrated value equal to zero. Conversely, in chiral configurations, the chirality density exhibits a spatially anisotropic distribution (shown in Fig. \ref{figure02}) and integrates to a non-zero finite value. Notably, the magnitude of the integrated chirality density is approximately equal for enantiomeric pairs, differing only in sign to reflect the handedness of the molecule. These observations confirm that chirality density is linked to electric toroidal multipoles, which are symmetry-allowed and active under the respective chiral point groups. These results underline the interplay between molecular symmetry, chirality, and the associated multipolar contributions, providing insights into how chiral phenomena emerge at a quantum mechanical level.


\subsubsection{Transmission studies}
\begin{figure}
\centering
\includegraphics[width=8.5cm,height=10.0cm]{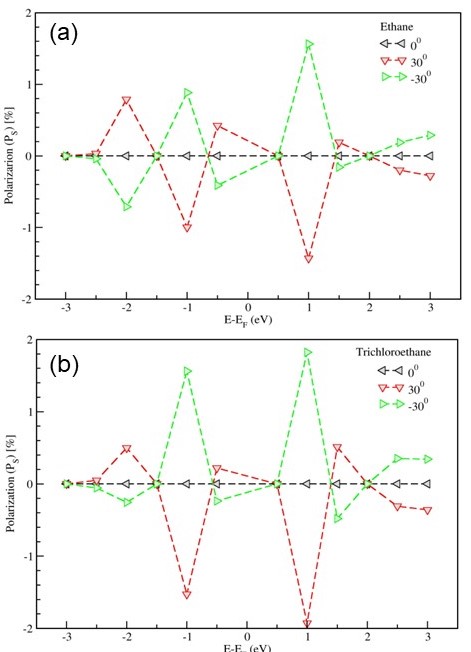}
 \caption{The spin polarization, $P_S$ of ethane (a) and trichloroethane (b) at zero bias, is presented as a function of energy, $E$.}
  \label{figure03}
\end{figure}

In order to connect  our observations on the chirality density more directly to the CISS effect using DFT-based electronic structure calculations, we applied the Landauer–Imry–Büttiker formalism \cite{Buttiker1993} to model transport as a coherent tunneling process (shown in Fig. \ref{figure03}). We evaluated, in particular, the (\textit{T}) were derived from LDA-based DFT calculations using the DIRAC code. In this study, we focus in particular on the isolated ethane molecules with chlorine substituents. The transport model used virtual hydrogen atoms as electrodes to approximate the local density of states (LDOS) of metallic leads, with the scattering region limited to the carbon and chlorine atoms. The results are presented here, which show both the spin polarization, $P_S$ and the extracted transmission function, $T$ for ethane and trichloroethane in dependence on the distance in energy from the Fermi level ($E_F$), comparing the results for different dihedral angles. This approach revealed that twisting the molecular dihedral angle modulates intrinsic SOC, providing information on the role of $P_S$ in CISS, a key factor in CISS as supported by previous studies \cite{Zollner2020,Dednam2023}. These results underscore the importance of developing robust DFT methods for accurate modeling of spin-selective transport.

\subsection{Static calculations with electric field}
\subsubsection{Chirality density}
\begin{figure}
\centering
\includegraphics[width=8.5cm,height=9.5cm]{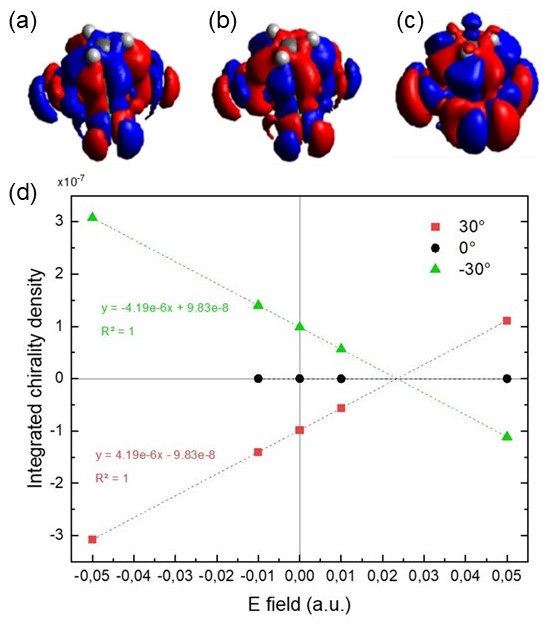}
 \caption{Perturbed chirality density in the $-30^{\circ}$ chiral trichloroethane molecule under external electric fields along the \textit{z}-axis with strengths of -0.01 (a), 0.01 (b), and 0.05 (c) atomic units. (d) The integrated chirality density as a function of the electric field strength. \textit{Note}: (i) the molecule's handedness determines the response's proportionality coefficient, and (ii) a positive bias (field opposing the molecular dipole moment) counteracts the inherent chirality density, flipping its sign above a critical field strength of \(\sim 0.0235\) atomic units.}
  \label{figure04}
\end{figure}

Chirality density corresponds to the expectation value of the helicity operator in the weak relativistic regime, both of the quantities showing sinusoidal dependence on the dihedral angle with confirmed proportionality ($R^2\sim0.99$). The proportionality constant between the integrated chirality density and helicity is approximately in agreement with theoretical predictions, being -137 for \textit{1,1,1-trichloroethane} (matching the atomic unit constant $mc$ used by \texttt{Dirac24}) and slightly higher (-163) for ethane, remaining consistent at all chiral angles. When an external static electric field is applied along the \textit{z} axis, the spatial distribution of the chirality density responds to the applied field (\textit{shown in} Fig. \ref{figure04} (a-c)). In particular, when we compute the perturbed spinor wavefunctions in the linear response regime, the integrated chirality density exhibits a linear dependence on the strength of the external field, as illustrated in Fig. \ref{figure04}(d). This behavior is compatible with a multipole interconversion mechanism \cite{kusunose2024} in the chiral $C_3$ conformations of \textit{1,1,1-trichloroethane}. A full-relativistic analysis might reveal an underlying SOC contribution of geometric origin.

\subsubsection{Transmission studies}
\begin{figure}
\centering
\includegraphics[width=8.5cm,height=10.0cm]{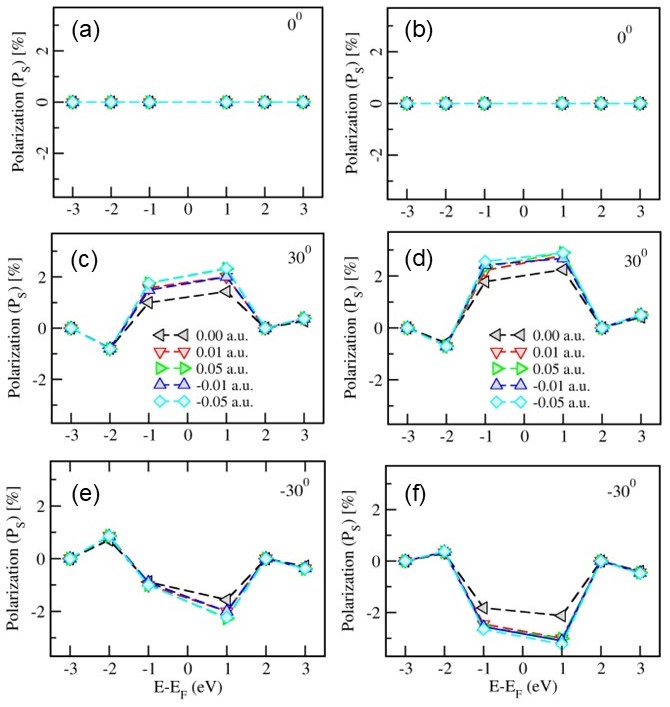}
 \caption{The spin polarization, $P_S$ of ethane at static finite bias of (a) $0^{\circ}$, (b) $30^{\circ}$ and (e) $-30^{\circ}$ with linear response, is presented as a function of energy. Similarly, (b) $0^{\circ}$, (d) $30^{\circ}$ and (f) $-30^{\circ}$ show for trichloroethane. The external electric field has a minimal impact. Similarly, the respective transmission function, \textit{T} could be extracted from this plot.}
  \label{figure05}
\end{figure}

CISS measurements are typically conducted under finite bias (typically resulting in an electric current), so the equilibrium zero bias approach used here may not capture all the relevant aspects of the phenomenon and that \textit{out-of-equilibrium} conditions might be essential for its full rationalization \cite{Zhi2021, Pan2020}. Although this method cannot fully simulate a nonequilibrium scenario, we approximate the electric potential within the junction by applying a linear electric field, using values of 0.01 a.u. and 0.05 a.u. in the forward direction and -0.01 a.u. and -0.05 a.u. in the reverse direction. The results (see Fig. \ref{figure05}(a-c) for C$H_3$-C$H_3$ and Fig. \ref{figure05}(d-f) for C$H_3$-C$Cl_3$) indicate that the linear field has a minimal impact (\textit{it is a few percent}) on $P_S$. Although the effect of the external electric field is small, this may change when the non-equilibrium condition is incorporated into the self-consistent solution of the electronic structure.

\section{Discussions}
The chirality density derived from ${\gamma^5}$, along with its integral, provides information on the cumulative chirality distribution throughout the system. The integrated chirality density exhibits sinusoidal dependence on the dihedral angle that defines geometric chirality in our systems, vanishing for achiral configurations and changing sign with handedness, as expected. Integrated chirality density also shows linear dependence on the external bias. In the absence of an external field, the proportionality between integrated chirality density and helicity matches theoretical predictions; however, finite fields reveal a field-dependent proportionality coefficient. These variations arise because helicity is less sensitive to the external field than chirality density, exhibiting oscillatory behavior without sign changes. The deviation may result from overlooked relativistic effects, contributions from the vector potential to kinetic momentum, or field-induced multipole couplings, potentially linking to Berry phase effects \cite{Abdiel2024,Brian2024}. It exhibits sinusoidal dependence, disappearing for achiral configurations and changing sign with handedness, as expected \cite{Dednam2023,Desmarais2024}. In the absence of an external field, their proportionality matches theoretical predictions; however, finite fields reveal a field-dependent proportionality coefficient. These findings indicate that SML, which originates from the intrinsic chirality density of electronic wavefunctions, plays a crucial role in CISS and its underlying mechanisms \cite{Hoshino2023, Sumit2023}. 

Using the relativistic DFT implementation within the \texttt{Dirac24} code, we have investigated the chirality of charge carriers by analyzing the density associated with the $\gamma^5$ matrix in Dirac's formalism. By studying molecular systems with tunable geometries, we observe that structural chirality induces a non-trivial, anisotropic distribution of particle chirality density that does not integrate to zero. Furthermore, the integral of the particle chirality is enhanced by the presence of dipoles or electric fields aligned with the chiral axis, suggesting the potential for generating spin-polarized currents in conduction experiments under non-equilibrium conditions \cite{Krieger2024,Shitade2020}. While these results are qualitatively consistent with the experimental evidence observed for this class of phenomena, the integral chirality we calculate in response to structural torsion is significantly smaller than the corresponding helicity values reported in experiments \cite{Wang2024,Zhang2023}. This discrepancy may arise from several factors, which we discuss below:

\textit{Non-Equilibrium Effects and Excited States-} Experimental conditions often involve non-equilibrium dynamics, where the mixing of the ground state with excited states, potentially exhibiting higher chirality, could play a critical role. Our calculations currently focus on equilibrium properties and may not fully capture these effects.

\textit{Incomplete Representation of SOC -} The SOC in its current atomic, single-particle formulation may be insufficient to account for the full phenomenology of the CISS effect. This points to the importance of exploring geometrical SOC contributions, which might descend from a chiral formulation of the Dirac equation, and electron-electron interactions, such as a dependence of the exchange-correlation functional (\( E_{xc} \)) on spin-current density within a \( U(1) \times SU(2) \) invariant framework.

\textit{Geometrical Effects and Multipole Formulations -} The role of geometrical factors, such as structural distortions, in inducing the unconventional electric toroidal multipole and possibly other higher-order multipole contributions to densities and fields, may need to be incorporated to better describe the interplay between chirality and spin polarization. These considerations highlight the need for more advanced theoretical frameworks and methodologies that can fully account for the interplay between geometry, SOC, and electron-electron interactions in chiral systems.

\section{Conclusions and Future Outlook}
We demonstrate that the expectation value of \(\gamma^5\) takes finite nonzero values in geometrically chiral systems. In the presence of an electric field, chiral molecules can exhibit regions at their ends with dominant \(\langle \gamma^5 \rangle\) distributions of opposite signs. We studied the transmission employing the Landauer-Imry-Büttiker formalism in presence of an electric field and found that the polarization is sensitive to the external field. We obtained results in good overall qualitative agreement with the experiment demonstrating a link between particle and geometrical chirality. Here, precise quantitative assessments of these effects will probably require more advanced computational tools and the need for more advanced exchange-correlation functional within the current density functional theory \cite{Desmarais2024}.

\section{Acknowledgements} 
This work is supported by the PRIN initiative of the Italian Ministry of Research and University (MUIR), under grant no. F53D23001070006-
CISS-PRIN22-2022FL4NZ4. \\

\textit{\textbf{Keywords}: 4C Dirac equation; Chirality-induced spin selectivity; full relativistic dynamics; Spin-orbit coupling; Electron transport; Isolated chiral molecules} 

\bibliography{apssamp}

\end{document}